\documentclass[reprint, superscriptaddress, amsmath, amssymb, aps, prx, longbibliography]{revtex4-2}
\usepackage{color}
\usepackage{hyperref}
\usepackage{graphicx}	
\usepackage{dcolumn}   	
\usepackage{bm}		    
\usepackage{epstopdf}
\usepackage{float}
\DeclareMathOperator{\Tr}{Tr}
\DeclareMathOperator{\dt}{dt}
\DeclareMathOperator{\dx}{dx}
\DeclareMathOperator{\dy}{dy}

\DeclareMathOperator{\x}{x}
\DeclareMathOperator{\dif}{d}
\DeclareMathOperator{\kp}{k}
\DeclareMathOperator{\rp}{r}
\DeclareMathOperator{\vp}{v}

\DeclareMathOperator{\q}{q}

\DeclareMathOperator{\dJp}{dJ}
\DeclareMathOperator{\Jp}{J}

\usepackage{array}
\newcolumntype{C}[1]{>{\centering\let\newline\\\arraybackslash\hspace{0pt}}m{#1}}

\begin{document}
\title{Steady-state theory of electron drag on polariton condensates}

\author{S. Mukherjee}
\affiliation{Department of Physics and Astronomy, University of Pittsburgh, Pittsburgh, PA 15260, USA}
\author{A. S. Bradley}
\affiliation{Jack Dodd Center for Quantum Technology, Department of Physics, University of Otago, Dunedin, New Zealand}
\author{D. W. Snoke}
\affiliation{Department of Physics and Astronomy, University of Pittsburgh, Pittsburgh, PA 15260, USA}

\begin{abstract}
We present a general theory of drag on a condensate due to interactions with a moving thermal bath of non-condensate particles, adapted from previous theory of equilibration of a condensate in a trap. This theory can be used to model the polariton drag effect observed previously, in which an electric current passing through a polariton condensate gives a measurable momentum transfer to the condensate, and an effective potential energy shift.   
\end{abstract}

\maketitle
\section{Introduction}

It is often stated in simplified terms that a superfluid does not experience drag. More accurately, a superfluid does not react to transverse forces, such as the sliding friction of a wall in a rotating ring condensate \cite{annett2004superconductivity,nozieres2018theory,lifshitz2013statistical}, and has quantized vorticity, so that a superfluid has no turbulence in the limit of low excitations. A superfluid will still react to longitudinal forces and body forces, such as the force of gravity. It is also well established from experiments and theory of cold atom condensates \cite{griffin2009bose,blakie2008dynamics,gardiner1998quantum,weiler2008spontaneous,neely2013characteristics,neely2010observation,bradley2008bose,bradley2015low,mcdonald2016brownian,rooney2010decay,gauthier2019giant} that a condensate out of equilibrium experiences damping and dissipation due to scattering of condensate particles with non-condensate particles. 

It is a natural consequence of this damping force due to non-condensate particles that a steady-state flow of non-condensate particles in one direction will lead to a net drag force on the condensate. Such a situation has already been demonstrated by our experimental work \cite{myers2018pushing} for a polariton condensate in a wire with a net flow of free electrons moving through the condensate in one direction. In those experiments, two effects were observed: first, the average momentum of the condensate was shifted up and down due to the collisions with the electrons, and second, under the same conditions, a shift of the chemical potential of the condensate was seen. One of the advantages of polariton condensate experiments is that both the momentum and energy of the condensate can be easily observed {\em in situ}, non-destructively, by spectroscopy of photons slowly leaking out of the condensate, which have a one-to-one mapping to the state of the polaritons from which they came \cite{Microcavities2017,deng2010exciton,carusotto2013quantum}.

In this paper, we adapt previous theory \cite{blakie2008dynamics,mcdonald2019anomalous} on the damping of nonequilibrium condensates to the case of an exciton-polariton condensate (henceforth simply referred to as polariton condensate), in which the reservoir of non-condensate particles is a fermionic electron gas instead of just the excited states of the same type of particle as in the condensate. This case is realistic for the polariton condensate experiments, in which the coherent condensate fraction was very high, 90\% or more, so that the dominant drag force came from free electrons driven through the system. We will assume that the density of the electrons is low enough, and their temperature high enough, that Pauli statistics do not come into play, and they can be modeled as a classical Boltzmannian gas. Although we will apply this theory to the specific case of a polariton condensate, the theory is quite general, applying to any condensate experiencing a net drag force. 

In system-reservoir theory for open quantum systems \cite{gardiner2004quantum,gardiner2015quantum}, energy damping in a system arises from its interactions with the reservoir which result in both particle and energy transfer. In far-from-equilibrium or non-steady-state scenarios, processes which do not conserve the number of particles in the condensate, e.g. collisions in which a condensate particle scatters with a non-condensate particle and leaves the condensate, going into the excited-state reservoir, are known to dominate over processes which conserve particle number in the condensate (and in the reservoir), e.g., a collision in which a condensate particle transfers energy to a particle in the reservoir, but remains in the condensate. This second process will occur even if the reservoir is a distinct species from the condensate, such as a fermionic electron gas. While energy damping is important in thermalization of a condensate out of equilibrium, this latter scenario is also important when a steady state is established between the system and the reservoir. In this paper we describe the process of energy damping via a two-particles-in, two-particles-out elastic collision between the particles of the system and the reservoir. In particular we emphasize on the connection of this process leading to a drag force acting on the system. 

Polaritons inherently have a finite lifetime $\sim 20 - 200$ ps inside a microcavity. It has been shown \cite{mitprl} that in the upper range of these lifetimes, polaritons can reach thermal equilibrium. In general, it is possible to observe the time evolution of a polariton condensate far from equilibrium all the way to equilibrium \cite{mukherjee2019observation,mukherjee2020dynamics,abbarchi2013macroscopic,lagoudakis2010coherent}; spectroscopy and imaging of the leaked photons from the microcavity give direct access to the polariton states inside the microcavity.  In this paper, we consider the case of a steady-state system with continuous generation and decay of the condensate, with constant unidirectional flow of the electron reservoir, which has constant density. 
 
It may at first seem surprising that polaritons, which are electrically neutral and hence do not respond to electric field, will respond at all to an electric current. A polariton spends part of its existence as an exciton, however, which is a bound electron-hole pair, and excitons interact with free electrons in the same way that a hydrogen atom interacts with a free electron, as a charged dipole interacting with a free charge, with a short-range potential \cite{hartwell, malpuech2003polariton, rapaport2000negatively, galbraith1996comparison}. Because of this, the polariton drag effect \cite{myers2018pushing} is effectively a new type of nonlinear process in which an electron transfers its momentum to a photon; in the entire ``life cycle'', a photon enters the system, is virtually absorbed into becoming an exciton; while it is in that state, it receives a momentum kick from a free electron; and then finally the exciton turns back into a photon which is emitted from the system with an altered momentum. In this paper, we will not consider in detail the dipole-free charge interaction, and will instead simply model the electron-polariton interaction as a short-range elastic collisional process.



\section{Theory}\label{Sec2}

Let us consider as an example a classical object moving through a viscous fluid. The drag force on the object is proportional to the relative velocity between the object and the fluid, and there is an equal but opposite force acting on the fluid. When the fluid is not moving, this drag force is responsible for dragging the fluid along with a moving object, and if the object is initially at rest, it can be moved by the drag force of a moving fluid. Our theory aims to capture this effect by showing that the case of a condensate colliding with electrons generates an effective, real-valued potential which depends on the relative motion of the condensate and electrons in the same way as in the case of a classical object moving through the viscous fluid. 

The interaction between the polariton condensate $\psi(\vec{r},t)$ and the electrons $\phi(\vec{r},t)$ which comprise the reservoir is modelled using hard core repulsive collisions. The interaction Hamiltonian is given by
\begin{equation}\label{eqn:H_int}
    H_{int}  = g\int d^2 r \Big(\psi^{\dagger}(\mathbf{r}) \phi^{\dagger}(\mathbf{r})  \phi(\mathbf{r}) \psi(\mathbf{r})\Big),
\end{equation}  
where the scalar fields $\psi(\vec{r},t)$ and $\phi(\vec{r},t)$ obey the equal time bosonic commutation and fermionic anti-commutation relations, respectively. The fields evolve under the Hamiltonian $H=H_\psi+H_\phi+H_{int}$, where $H_\psi$ and $H_\phi$ are given by
\begin{equation}\label{eqn:Hpsi&phi}
\begin{aligned}
& {H}_\psi =\int d^2r \psi^\dagger(\mathbf{r})\Bigg[\frac{1}{2m_\psi}(-i\hbar\vec{\nabla}-\vec{p}_{rel})^2+V_\psi(\mathbf{r})\\ 
& \ \ \ \ \ \ +g n_\phi(\mathbf{r})+\frac{U}{2}\psi^\dagger(\mathbf{r})\psi(\mathbf{r})\Bigg]\psi(\mathbf{r}), \\
& H_\phi= \int d^2r \phi^\dagger(\mathbf{r})\Bigg[\frac{1}{2m_\phi}(-i\hbar\vec{\nabla})^2\Bigg]\phi(\mathbf{r}).
\end{aligned}
\end{equation}
Here $g n_\phi(\boldsymbol{r})$ is the Hartree energy shift due to the interactions between the electron reservoir and the condensate, and $U$ is the strength of the polariton-polariton repulsive interactions. The momentum $\vec{p}_{rel} = m_\psi\mathbf{v_{rel}} = m_\psi(\mathbf{v_\psi}-\mathbf{p_\phi}/m_\phi )$, is the steady-state drift momentum of the condensate measured in the rest frame of the electron reservoir. The electron momentum $\mathbf{p_\phi}$ is the steady-state average linear momentum of the reservoir measured in the lab frame due to the application of an electric field on the electron reservoir. The electrons in the quantum well are assumed to be freely moving and the effect of applying a constant potential in the plane of the well is taken as setting up a steady-state flow with a drift momentum $\mathbf{p_\phi}$. $V_\psi(\mathbf{r})$ is the effective potential experienced by the polaritons due to the photonic energy gradient in the microcavity and repulsive potential due to the interaction with the excitons created by the non-resonant pump.  

Our goal in this section is to derive an effective equation of motion for the polariton condensate, ignoring both the thermal and quantum fluctuations of the condensate order parameter. We adopt the formalism of the Master equation approach which has been applied previously to the theory of finite temperature Bose-Einstein condensates to describe number and energy dissipation in such systems \cite{blakie2008dynamics,gardiner2004quantum}. An important consideration in this approach is that the states comprising the system and the reservoir have different energy scales; low-lying energy states make up the ``system'' while high-energy excited states make up the ``reservoir''. Such a distinction introduces a cut-off energy parameter separating the two regions (coherent modes of the system from the incoherent, thermal modes of the reservoir), which finds its way into the scattering rate and at first glance seems ad-hoc. In the theory of stochastic Gross-Pitaevskii equations, this cut-off parameter is chosen self consistently using Hartree-Fock theory, conserving the total number of particles, and is found to predict quantitative estimates for observables measured in the experiments \cite{bradley2008bose,neely2010observation,gauthier2019giant}. In the case considered here, of a fermionic electron reservoir, there is no need for such an energy cutoff because the reservoir particles are a different species, completely distinguishable from the condensate particles.

We assume that the incoherent reservoir maintains thermal equilibrium while interacting with the system, due to strong interactions with lattice phonons. The polaritons, on the other hand, have weak coupling with the phonons \cite{hartwell}, and thus must come to equilibrium via interactions with each other, with excitons, and with the free electrons. We assume that the electrons have a much higher average kinetic energy than the polaritons, because the polariton condensate fraction is so high that it mostly occupies very low energy states. 

The starting point of the theory is to describe the time evolution of the system density matrix $\rho$ (including polaritons and the reservoir) under the full Hamiltonian $H$, as follows:
\begin{equation}\label{eqn:rho_t}
    \partial_t\rho(t) = -\frac{i}{\hbar}[H,\rho(t)].
\end{equation}
In the interaction picture, $H_{int}(t) = e^{i(H_\psi+H_{\phi})t/\hbar}H_{int}e^{-i(H_\psi+H_{\phi})t/\hbar}$ and the interaction picture density matrix $\rho_I(t)=e^{i(H_\psi+H_{\phi})t/\hbar}\rho(t)e^{-i(H_\psi+H_{\phi})t/\hbar}$, which implies
\begin{equation}\label{eqn:rho_int}
    \partial_t\rho_I(t)=-\frac{i}{\hbar}\Big[H_{int}(t),\rho_I(t)\Big].
\end{equation}
\begin{widetext}
The above equation can also be written as
\begin{equation}
     \partial_t\rho_I(t)=-\frac{i}{\hbar}\Big[H_{int}(t),\rho_I(0)\Big]\\
     -\frac{1}{\hbar^2}\int_{0}^{t} dt' \Bigg[H_{int}(t),\Big[H_{int}(t'),\rho_I(t')\Big]\Bigg].
\end{equation}
Taking the partial trace $Tr_\phi[...]$ over the electron field leads to

\begin{equation}\label{eqn:MasterEqn}
    \partial_t Tr_{\phi}\Bigg[\rho_I(t)\Bigg]=Tr_{\phi}\Bigg[-\frac{i}{\hbar}\Big[H_{int}(t),\rho_I(0)\Big]\\
    -\frac{1}{\hbar^2}\int_{0}^{t} dt' \Bigg[H_{int}(t),\Big[H_{int}(t'),\rho_I(t')\Big]\Bigg]\Bigg].
\end{equation}
\end{widetext}

We assume that at the initial time the electrons and the polaritons are uncorrelated, allowing us to represent the density matrix at $t=0$ as a direct product of the two subsystems $\rho_I(0)=\rho_\psi\otimes\rho_\phi$. The first term on the right hand side of the above equation gives a Hartree energy contribution = $g n_\phi(\mathbf{r})$ which has already been absorbed in the definition of $H_\psi$. Therefore this term is dropped. Further, we assume that the electron-polariton interactions are weak, so that correlations between the reservoir and the polaritons are small even on long time scales, such that $\rho_I(t')\approx \rho^I_\psi(t')\otimes\rho_\phi$. As discussed above, we assume that the energy separation between the electrons and the polaritons is large, which implies a much slower evolution of $\rho^I_\psi(t')$, allowing us to replace $\rho^I_\psi(t')\rightarrow\rho^I_\psi(t)$ in the time integral. This approximation is known as the Markov approximation and amounts to assuming that the evolution of the density matrix depends on the instantaneous state of the system and makes no reference to the past. 
After evaluating the bath correlation functions we arrive at
\begin{widetext}
\begin{equation}\label{eqn:densityMatrix1}
  \begin{aligned}
   &\partial_t\rho_\psi(t) = -\frac{i}{\hbar}\Big[{H}_\psi,\rho_\psi(t)\Big] -\frac{\pi \hbar g^2}{4\pi^2\left(2\pi\hbar\right)^2} \int d^2r d^2r' d^2k_1 d^2k_2 F(\boldsymbol{k_1})\Big(1-F(\boldsymbol{k_2})\Big) \\ 
 & \Bigg\{ \Big[\hat{n}_\psi(\boldsymbol{r}), \delta\left(\hat{L}_\psi+E_\phi(\boldsymbol{k_1})-E_\phi(\boldsymbol{k_2})\right)\hat{n}_\psi(\boldsymbol{r}')\rho_\psi(t)\Big]e^{-i(\boldsymbol{k_1}-\boldsymbol{k_2}).(\boldsymbol{r}-\boldsymbol{r}')}\\
& +\Big[\rho_\psi(t)\delta\left(\hat{L}_\psi-E_\phi(\boldsymbol{k_1})+E_\phi(\boldsymbol{k_2})\right)\hat{n}_\psi(\boldsymbol{r}'), \hat{n}_\psi(\boldsymbol{r}) \Big]e^{i(\boldsymbol{k_1}-\boldsymbol{k_2}).(\boldsymbol{r}-\boldsymbol{r}')}\Bigg\},
  \end{aligned}
\end{equation}
\end{widetext}
with the definitions $\rho_{\psi}(t) = Tr_{\phi}\Big[\rho(t)\Big]$, $\hat{n}_\psi(\boldsymbol{r})=\psi^\dagger(\boldsymbol{r})\psi(\boldsymbol{r})$, $\hat{L}_\psi\hat{n}_\psi(\boldsymbol{r})= [\hat{n}_\psi(\boldsymbol{r}),H_\psi]$, $E_\phi(\boldsymbol{k})=\hbar^2\boldsymbol{k}^2/2m_\phi$ and $F(\boldsymbol{k})=1/(e^{(E_\phi(\boldsymbol{k})-\mu)/k_B T}+1)$, assuming the electronic reservoir is in thermal equilibrium with chemical potential $\mu$ and temperature $T$. Details of the calculation in going from \eqref{eqn:MasterEqn} to \eqref{eqn:densityMatrix1} are given in Appendix \ref{appendix:Appendix_derivation}. As shown in Appendix \ref{appendix:Appendix_derivation}, the fermionic nature of the reservoir is manifested in the above expression as the factor $F(\boldsymbol{k_1})\Big(1-F(\boldsymbol{k_2})\Big)$, which reduces to Maxwell-Boltzmann distribution in the high temperature and low electron density limit as shown in Appendix \ref{appendix:Appendix_scattering}. For book-keeping purposes, we define the eigenenergies of the super-operator $\hat{L}_\psi$, which are given by $\{\epsilon\}$. With this we define
\begin{widetext}
\begin{equation}\label{eqn:scatteringMatrix1}
\begin{aligned}
   & M_1(\boldsymbol{r}-\boldsymbol{r}',\epsilon) = \int d^2k_1 d^2k_2 F(\boldsymbol{k_1})\Big(1-F(\boldsymbol{k_2})\Big)\delta\left(\epsilon+E_\phi(\boldsymbol{k_1})-E_\phi(\boldsymbol{k_2})\right)e^{-i(\boldsymbol{k_1}-\boldsymbol{k_2}).(\boldsymbol{r}-\boldsymbol{r}')},\\
   & M_2(\boldsymbol{r}-\boldsymbol{r}',\epsilon) = \int d^2k_1 d^2k_2 F(\boldsymbol{k_1})\Big(1-F(\boldsymbol{k_2})\Big)\delta\left(-\epsilon+E_\phi(\boldsymbol{k_1})-E_\phi(\boldsymbol{k_2})\right)e^{i(\boldsymbol{k_1}-\boldsymbol{k_2}).(\boldsymbol{r}-\boldsymbol{r}')}.
   \end{aligned}
\end{equation}
\end{widetext}
Interchanging $\boldsymbol{k_1}$ and $\boldsymbol{k_2}$ in $M_1(\boldsymbol{r}-\boldsymbol{r}',\epsilon)$ and using the property $F(\boldsymbol{k_2})\Big(1-F(\boldsymbol{k_1})\Big)=e^{\beta\epsilon}F(\boldsymbol{k_1})$ $\Big(1-F(\boldsymbol{k_2})\Big)$ we arrive at,
\begin{equation}\label{eqn:forwardBackward}
    \frac{M_1(\boldsymbol{r}-\boldsymbol{r}',\epsilon)}{e^{\beta\epsilon/2}} = \frac{M_2(\boldsymbol{r}-\boldsymbol{r}',\epsilon)}{e^{-\beta\epsilon/2}}.
\end{equation}
For $\beta\epsilon\ll1$, we can linearize \eqref{eqn:scatteringMatrix1} using \eqref{eqn:forwardBackward} to obtain
\begin{equation}\label{eqn:forwardBackward2}
    \begin{aligned}
        & M_1(\boldsymbol{r}-\boldsymbol{r}',\epsilon)\approx\Big(1+\frac{\beta\epsilon}{2}\Big)M(\boldsymbol{r}-\boldsymbol{r}',0),\\
        & M_2(\boldsymbol{r}-\boldsymbol{r}',\epsilon)\approx\Big(1-\frac{\beta\epsilon}{2}\Big)M(\boldsymbol{r}-\boldsymbol{r}',0).
    \end{aligned}
\end{equation}
Using \eqref{eqn:scatteringMatrix1}, \eqref{eqn:forwardBackward} and \eqref{eqn:forwardBackward2}, we can approximate \eqref{eqn:densityMatrix1} as
\begin{widetext}
\begin{equation}\label{eqn:densityMatrix2}
  \begin{aligned}
   \partial_t\rho_\psi(t) \approx & -\frac{i}{\hbar}\Big[{H}_\psi,\rho_\psi(t)\Big] -\frac{\pi \hbar g^2}{4\pi^2\left(2\pi\hbar\right)^2} \int d^2r d^2r' 
 M(\boldsymbol{r}-\boldsymbol{r}')\Bigg\{\Big[\hat{n}_\psi(\boldsymbol{r}),\hat{n}_\psi(\boldsymbol{r}')\rho_\psi(t)\Big] \\ 
& + \Big[\rho_\psi(t)\hat{n}_\psi(\boldsymbol{r}'), \hat{n}_\psi(\boldsymbol{r}) \Big]  +\frac{\beta}{2}\Big[\hat{n}_\psi(\boldsymbol{r}),\hat{L}_\psi\hat{n}_\psi(\boldsymbol{r}')\rho_\psi(t)\Big]
-\frac{\beta}{2}\Big[\rho_\psi(t)\hat{L}_\psi\hat{n}_\psi(\boldsymbol{r}'), \hat{n}_\psi(\boldsymbol{r}) \Big]\Bigg\}.
  \end{aligned}
\end{equation}
\end{widetext}
Next, we map this equation to the Fokker-Planck equation in the semi-classical limit, which is found by introducing a Wigner-Weyl transform \cite{gardiner2004quantum,polkovnikov2010phase} of the density and field operators. The terms $\Big[\hat{n}_\psi(\boldsymbol{r}),\hat{n}_\psi(\boldsymbol{r}')\rho_\psi(t)\Big]$ and $\Big[\rho_\psi(t)\hat{n}_\psi(\boldsymbol{r}'), \hat{n}_\psi(\boldsymbol{r}) \Big]$ inside the integral map to second-order field derivatives, which give rise to a multiplicative noise in the time evolution of the c-field $\psi(\mathbf{r},t)$ \cite{gardiner2003stochastic}. As we are only interested in the coherent field dynamics, we drop this term. The terms proportional to $\beta$ transform to 
\begin{widetext}
\begin{equation}
\begin{aligned}
    & \frac{i\hbar \beta}{2}\vec{\nabla}'\cdot\bigg(\frac{\boldsymbol{p}_{rel}}{m_\psi} |\psi(\boldsymbol{r}')|^2 + \frac{i\hbar}{2m_\psi}\psi^*(\boldsymbol{r}')\vec{\nabla}'\psi(\boldsymbol{r}')-\frac{i\hbar}{2m_\psi}\psi(\boldsymbol{r}')\vec{\nabla}'\psi^*(\boldsymbol{r}')\bigg)\\
    & \bigg(\psi^*(\boldsymbol{r})\frac{\delta W}{\delta\psi^*(\boldsymbol{r})}-\psi(\boldsymbol{r})\frac{\delta W}{\delta\psi(\boldsymbol{r})}\bigg) \\
    & = -\frac{i\hbar \beta}{2}\vec{\nabla}'\cdot\bigg(\frac{\vec{p}_{rel}}{m_\psi} |\psi(\boldsymbol{r}')|^2 + \mathbf{j}(\mathbf{r}')\bigg)\bigg(\psi(\boldsymbol{r})\frac{\delta W}{\delta\psi(\boldsymbol{r})}-\psi^*(\boldsymbol{r})\frac{\delta W}{\delta\psi^*(\boldsymbol{r})}\bigg).
    \end{aligned}
\end{equation}
\end{widetext}
Here $W(\psi,\psi^*)$ is the Wigner function which is defined as the Weyl transform of the density matrix, $\nabla'$ denotes spatial derivatives with respect to $\mathbf{r}'$ co-ordinates, and $\mathbf{j}(\mathbf{r}')$ is the condensate current. Details of this mapping are discussed in Appendix \ref{appendix:Appendix_transform}. Similarly, applying the Wigner-Weyl transform to the other terms in \eqref{eqn:densityMatrix2}, we arrive at a semi-classical equation for the evolution of the Wigner function after ignoring the second and higher order field derivatives:
\begin{widetext}
\begin{equation}\label{eqn:densityMatrix3}
\begin{aligned}
    & \partial_t W = -\frac{i}{\hbar}\int d^2r \Big(L_{op} + V_\epsilon(\mathbf{r})\Big)\bigg(\psi(\boldsymbol{r})\frac{\delta W}{\delta\psi(\boldsymbol{r})}-\psi^*(\boldsymbol{r})\frac{\delta W}{\delta\psi^*(\boldsymbol{r})}\bigg),\\
    & L_{op} = \frac{1}{2m_\psi}(-i\hbar\vec{\nabla}-\vec{p}_{rel})^2+V_\psi(\mathbf{r})+g n_\phi(\mathbf{r})+U|\psi(\mathbf{r})|^2, \\
    & V_\epsilon(\mathbf{r}) = -\frac{\hbar\beta g^2}{32\pi^3}\int d^2r'M(\mathbf{r}-\mathbf{r}')\vec{\nabla}'\cdot\underbrace{\bigg(\frac{\vec{p}_{rel}}{m_\psi} |\psi(\boldsymbol{r}')|^2 + \mathbf{j}(\mathbf{r}')\bigg)}_{\mathbf{J}(\mathbf{r'})}.
\end{aligned}
\end{equation}
\end{widetext}
$V_\epsilon(\mathbf{r})$ is the field-dependent drag potential which is nonlinear. It is straightforward to show that the drag potential causes energy dissipation of the condensate, since it is proportional to the negative of the divergence of the condensate current. Since damping is a weak correction to Hamiltonian dynamics, at leading order $\nabla\cdot \mathbf{j}=-\partial_t |\psi|^2$, and the potential damps density dynamics associated with sound, solitons, or moving vortex cores \cite{mcdonald2019anomalous,robEhrenfest}. Note that a similar expression for the total current $\mathbf{\Jp}(\mathbf{r}')$ is obtained for a rotating condensate \cite{bradley2008bose}. The above equation describing the evolution of the Wigner function of the polariton condensate can be mapped to a noise-free equation for the evolution of the condensate order parameter \cite{gardiner2003stochastic}. To apply this theory to the interpretation of the experimental observations under consideration here, we first impose the geometrical constraints of the experiments. We assume that the condensate can flow only along one-dimension ($x$-direction) while its motion is frozen in the transverse direction ($y$-direction). The condensate is assumed to be in the ground state of the particle in a box potential of width $w$ along the $y$-direction. The condensate c-field can now be written as $\psi(x,y,t)= \sqrt{(2/w)}\cos{(\pi y/w)} \psi(x,t)$. The equation of motion for the condensate field $\psi(\mathbf{r},t)$ in one-dimension is given by integrating over the transverse dimension to give
\begin{widetext}
\begin{equation}
\begin{aligned}
   & i\hbar\partial_t\psi(x,t)  = \frac{2}{w}\int_{-w/2}^{w/2} dy \cos{(\pi y/w)} \big(L_{op}+V_\epsilon(x,y,t)\big)\cos{(\pi y/w)} \psi(x,t)\\
    & = \bigg(E_c + \frac{1}{2m_\psi}(-i\hbar\partial_x-p_{rel})^2+V_\psi(x)+g n_\phi(x)+\frac{3}{2w}U|\psi(x)|^2+V_\epsilon(x,t)\bigg)\psi(x,t).
\end{aligned}
\label{eqn:drag1}
\end{equation}
\end{widetext}
Here $E_c$ is the confinement energy due to the confinement along the y-direction and can be dropped. An expression for $V_\epsilon(x,t)$ is derived in Appendix \ref{appendix:Appendix_drag1D}. In a typical experiment, the electron density concentration $n_\phi$ can only be estimated and cannot be determined precisely, nor the chemical potential $\mu$ and the effective temperature $1/\beta$ of the electron reservoir. These parameters are free parameters in our theory, which we can adjust to obtain qualitative agreement with the overall behavior in the experiments, to show that the drag effect exists.

    \begin{figure*}
		\centering
		\includegraphics[width=0.8\textwidth]{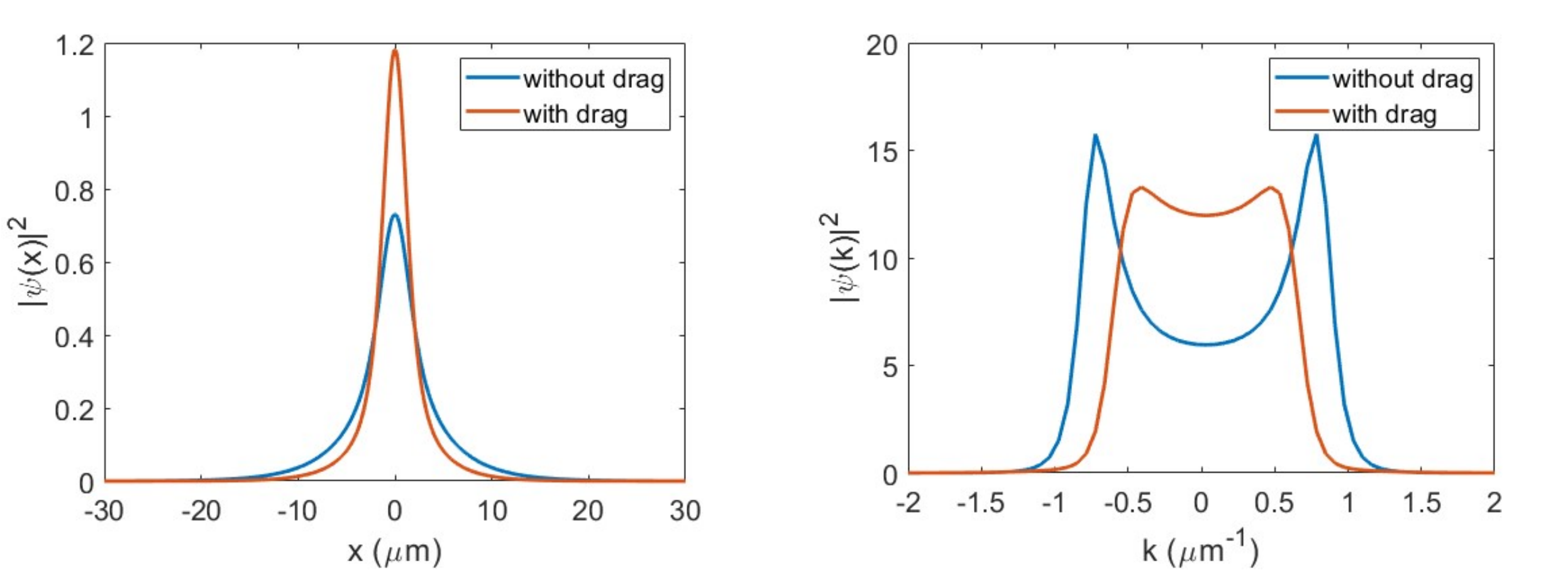}
		\caption[Drag effect on expanding wavepacket]{Example of the effect of the drag potential on a freely expanding polariton wavepacket in presence of a continuous generation and decay of polaritons. The left figure compares the spatial profile of the wavepacket in presence and absence of the drag potential when a steady state has been reached. The right figure shows the momentum distribution of the steady state corresponding to the spatial profiles shown left.}
		\label{fig:dragSimple}
	\end{figure*}

The Gross-Pitaevsksii equation for the polariton condensate is further adjusted by introducing a finite lifetime and decay out as photons from the microcavity, and generation of the polaritons by a source term proportional to an external pump laser intensity. The system can reach steady state when a continuous wave (CW) laser is tuned in wavelength to create excitons and exciton-polaritons at high energy, which then cool down into the condensate by stimulated scattering \cite{snoke89,snoke1994density,snoke2020solid}.

	\begin{figure*}
		\centering
		\includegraphics[width=0.5\textwidth]{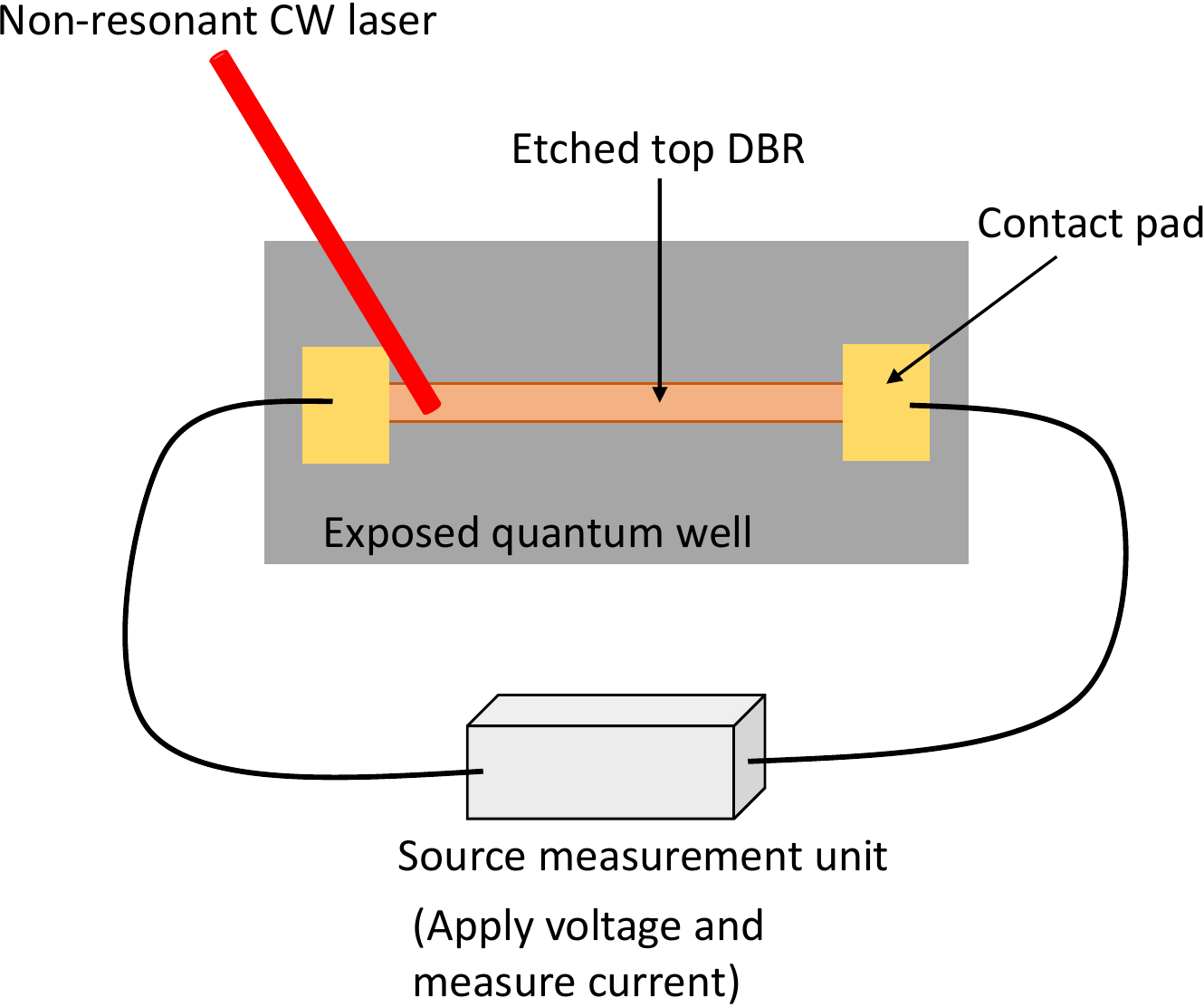}
		\caption[Sketch of the polariton drag experiment]{Top view sketch of the polariton wire fabricated by etching the top DBR layers. Polariton condensate is created in the wire by pumping non-resonantly with a CW laser near one of the edges of the wire. A current is driven through the wire by applying voltages at the two ends of the wire as shown above. }
		\label{fig:dragScheme}
	\end{figure*}



The rate of generation of the polaritons by the pump $P(x,t)$ and the decay of the polaritons from the microcavity as photons at a rate $\gamma$ is included in the continuity relation for the condensate density $n(x,t) = |\psi(x,t)|^2$ as
\begin{equation}
    \frac{dn(x,t)}{\dt} = P(x,t) - \gamma n(x,t) - \frac{\dJp (x,t)}{\dx},
    \label{eqn:cont1}
\end{equation}
where $\Jp (x,t)$ is the condensate current defined in Equation \ref{eqn:densityMatrix3}. Equations \ref{eqn:drag1} and \ref{eqn:cont1} describe an energy dissipative system with the total particle number given by the balance between the generation and loss of polaritons. These equations are evolved for a long time until a steady state is reached under a CW pump. 

Let us first consider a simple model to illustrate the effect of the drag potential on the condensate. By assuming very short polariton lifetime (2 ps) in a flat one dimensional wire we restrict the polaritons closer to the region of their generation because their population decays rapidly. They are produced on top of a ``hill" (localized potential energy maximum) due to the repulsion of the polaritons with the excitons at the pump location. When there is no drag potential, the expansion of the condensate is given by the kinetic and the repulsive polariton-polariton interactions as well as the rate of generation and decay of the polaritons. This creates a population of the condensate extending outside the pump region. When the drag potential is introduced, the condensate density looks squeezed with more particles in the center than in the tails of the spatial profile, as shown in Figure \ref{fig:dragSimple}. In Figure \ref{fig:dragSimple} we also see that the momentum distribution is altered when the drag potential is turned on. The drag potential slows down the velocity of the particles moving outwards as a result the maximum velocity reached by the particles is smaller when drag is present. A smaller velocity distribution of the particles imply that the particles cannot travel as far in their lifetime as when the drag is absent. This results in a shrunken tail of the condensate spatial distribution as seen in Figure \ref{fig:dragSimple}. Another aspect of the drag potential is discussed in the next section, which gives an energy shift to the chemical potential of the condensate.          

\section{Polariton drag effect}\label{Sec3}

\begin{figure*}
		\centering
		\includegraphics[width=\textwidth]{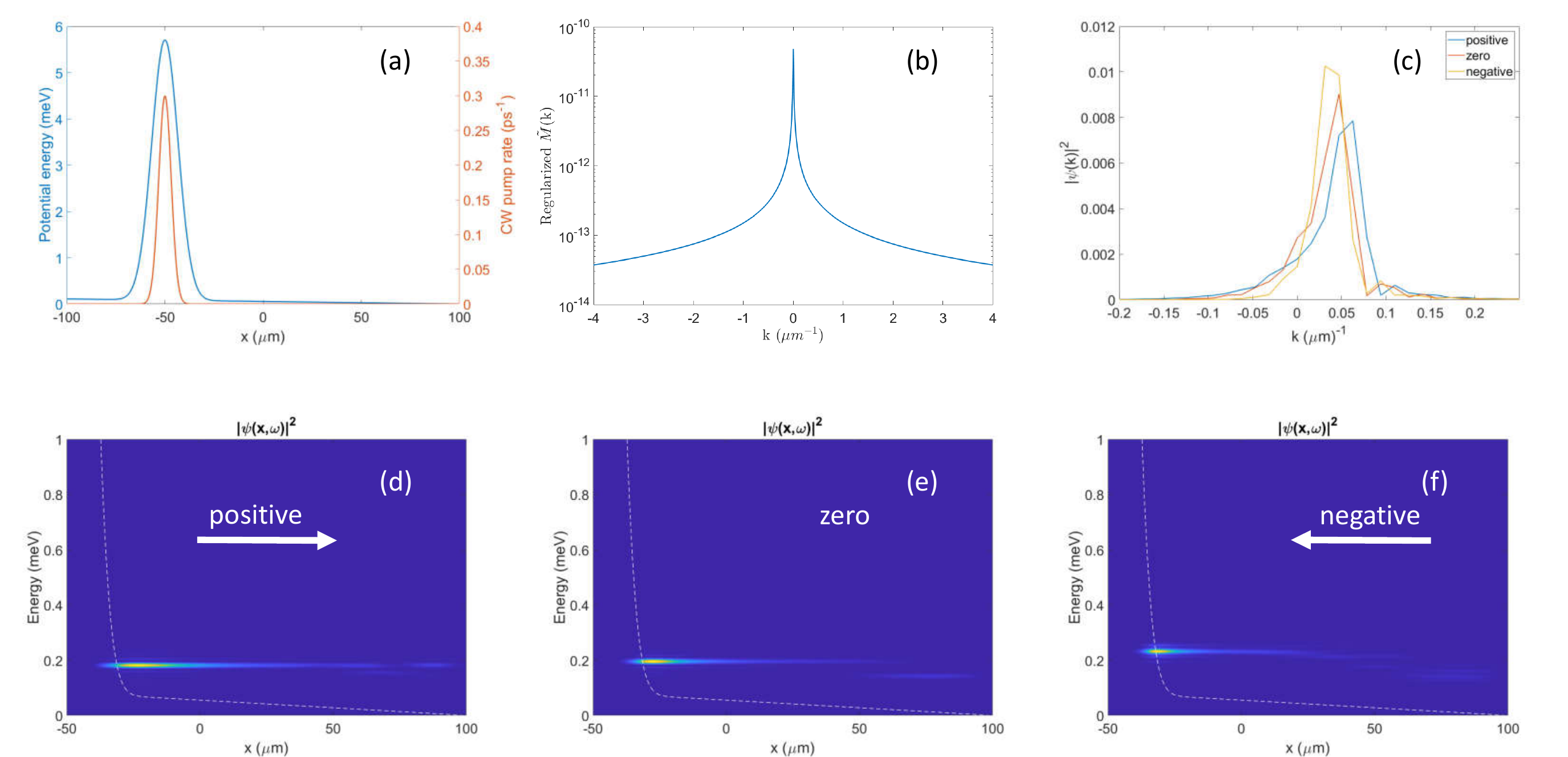}
		\caption[Comparison of the drag effect]{(a) The orange curve shows the pump which acts as a source generating the polaritons in the wire. The blue curve shows the effective potential experienced by the polaritons in the wire. (b) Regularized scattering amplitude used in the numerical simulations is shown here. (c) Time-integrated average momentum of the condensate corresponding to the energy spectra in (d-f) is shown here. (d-f) Time-integrated spectra of the condensate along the wire for different relative velocities of the electron reservoir. The white dashed line serves as a guide to the eye for observing the changes in the condensate energy. Real space spectra is obtained by the Fourier transform of the late time wavefunction in a temporal window and averaged over many such windows. In the simulation, the parameters used were $m_\psi=1\times10^{-4}m_e$, $U=10 \ \mu eV \mu m^2$, $v_{rel}=2\times 10^5$ m/s.   Each image is plotted on a normalized color scale from 0 - 1.  }
		\label{fig:dragScheme2}
	\end{figure*}

In this section we discuss the qualitative features observed in the experiment given in reference \cite{myers2018pushing}. A simple schematic of the experimental setup is shown in Figure \ref{fig:dragScheme}. In the experiment, a tightly-focused pump spot generated polaritons near one end, but not exactly at the end, of a one-dimensional (1D) wire. Free excitons generated by the laser (in addition to the condensate polaritons) created a potential energy maximum felt by the polaritons at that point, since the excitons have a strong repulsion on the polaritons. Some of these excitons diffuse away from the pump spot \cite{MyersArxiv2018}, giving a smeared-out spatial peak that repels the polariton condensate. An example of an effective potential experienced by the polaritons in the wire, used in the numerical model, is shown in Figure \ref{fig:dragScheme2}(a) blue curve, and the spatial profile of the pump generating the polariton condensate is shown in orange curve. 

The polariton condensate flows away from this generation spot in both directions along the 1D wire. At the end of the wire nearest to the pump spot, a local trap is formed for the condensate, in which the condensate velocity is nearly zero; on the other side, which has much longer distance for the condensate to travel, a steady-state flow is set up in which the polaritons have net velocity away from the pump spot, as they are continuously generated at the pump spot and then decay away as they travel. Angle-resolved spectroscopy experimentally determined the average momentum of this moving condensate \cite{myers2018pushing}. Electric current was introduced into the same structure in which the optically-generated condensate flowed, using electrical contacts. The energy of both the extended as well as the localized condensate were found to shift in response to the magnitude and direction of the electric current. In those experiments, as no energy shift was observed at the location of the pump spot, which is primarily composed of excitons, any excitonic mechanism contributing to these energy shifts due to the variation of the exciton density could be ruled out. 

The drag potential dissipates energy of the condensate resulting in the formation of a low energy steady state condensate in the wire. This is shown in the real-space, time-integrated spectra of the condensate moving to the right (positive x-direction) in the wire in Figures \ref{fig:dragScheme2}(d-f). In these simulations we assumed a long polariton lifetime (200 ps) consistent with the long polariton samples used in the recent experiments \cite{steger2013long,steger-turn,myers2018pushing,mukherjeePRB2021,mukherjee2019observation}. The white dashed line in this figure outlines the potential energy felt by the polaritons in the wire. The low energy steady state is reached after an evolution of 4 ns from a high energy initial state. We used no additional imaginary terms in the simulation to observe this energy dissipation. The real valued drag potential causes the energy dissipation in the system. The momentum-dependent scattering amplitude used in the simulation is shown in Figure \ref{fig:dragScheme2}(b) as derived in Appendix \ref{appendix:Appendix_scattering}. Since the condensate is moving to the right, the net momentum is non-zero and positive as shown in the Figure \ref{fig:dragScheme2}(c) with no applied voltage. When a negative voltage is applied, the moving electrons apply a force on the condensate to the left, which slows down the condensate. Conversely, when a positive voltage is applied the condensate receives a kick from the moving electron reservoir and the net momentum is increased. These features are shown in Figure \ref{fig:dragScheme2}(c). We now compare real space spectra of the condensate when the reservoir is moving. The center of mass of the condensate is shifted either to the left or right when the reservoir is in motion relative to the stationary reservoir case. This is expected since the reservoir applies a body force on the condensate, effectively dragging the condensate with it. This is consistent with the momentum change as seen in Figure \ref{fig:dragScheme2}(c). In addition, from these spectra we observe energy shifts of the condensate when the reservoir is moving. These signatures when considered together confirm the polariton drag effect observed in the experiments \cite{myers2018pushing}. In these experiments, the electron concentration can be considered homogeneous in the quantum well, therefore it will result only in a uniform blue shift of the energy of the condensate due to polariton-electron interaction. As the photoluminiscence intensity, which directly indicates the density of the condensate, was not found to be significantly different for different applied voltages, we can rule out the shifts in the condensate energy due to the polariton-polariton and polariton-exciton interactions. What remains is the effect of the real-valued effective potential derived for the polariton drag in the previous section of this paper.

In a closed system with no particle generation or loss, the drag potential will cause any excited state of a trapping potential to evolve towards the stationary ground state of the trap. Once this state is reached, the drag potential becomes zero since there is no net current in this state. This scenario  is different since the condensate in the trap does not reach the stationary ground state of the trap, instead it reaches a non-equilibrium steady state. In this state, the drag potential is non-zero since there is a finite inhomogeneous particle current in the condensate. By creating a moving reservoir we probe the change in the drag potential which gives an energy shift to the condensate. This is clearly seen in Figures \ref{fig:dragScheme2}(d-f). There is relative shift of the energy of the condensate with respect to the stationary reservoir scenario when the reservoir is either co-moving with the condensate or moving in the opposite direction. Moreover, the shift in the energy is not symmetric for the motion of the electron reservoir in the same and the opposite directions to the motion of the condensate. This asymmetric shift could be understood as due to the different velocities seen by the condensate in the rest frame of the reservoir, when the reservoir moves with or against the condensate.    

\section{Conclusion}\label{Sec4}

We have shown that the general theory for dissipation of a nonequilibrium atomic Bose-Einstein condensate can be directly adapted to describe the case of a steady-state condensate with net current of a different species passing through it. This theory directly reproduces the experimental results for polariton drag, that is, the direct effect of DC electric current on the velocity and energy of neutral polaritons. The same theory underlies the cooling of a nonequilibrium condensate, which has been observed experimentally in a ring geometry \cite{mukherjee2019observation}. 

The general picture that emerges from this theoretical model is one in which the condensate keeps its overall coherence, and is well described by a modified Gross-Pitaevskii equation for a single-valued wave function, but still undergoes energy damping and drift force. The drag potential which has been the subject of this paper is shown to originate microscopically from the number conserving interactions of the polariton condensate with the incoherent electron reservoir leading to energy dissipation of the condensate. More common way of introducing energy damping is through an imaginary term \cite{choi1998phenomenological} which also arise from the same underlying theory of interaction of the condensate with an incoherent reservoir but leads to number damping and adapted in the previous papers for modeling energy loss in the system.

{\bf Acknowledgements}. SM thanks Daniel Boyanovsky for useful discussions. This work was supported by the National Science Foundation under Grant No. DMR-2004570. AB acknowledges support from the Marsden Fund with grant No. UOO1726, and the Dodd-Walls Centre for Photonic and Quantum Technologies. This research was supported in part by the University of Pittsburgh Center for Research Computing through the resources provided.

\bibliography{etd}

\appendix

\section{Quantum Master equation}\label{appendix:Appendix_derivation}
In this appendix we will fill the steps between \eqref{eqn:MasterEqn} and \eqref{eqn:densityMatrix1}. As mentioned earlier, we'll assume that at the initial time the electrons and the polaritons are uncorrelated implying that we could represent the density matrix at $t=0$ as a direct product of the two subsystems $\rho_I(0)=\rho_\psi\otimes\rho_\phi$. With this let us now unpack the first term of \eqref{eqn:MasterEqn},
\begin{widetext}
\begin{equation}
    \Tr_{\phi}\Bigg[-\frac{i}{\hbar}\Big[H_{int}(t),\rho_I(0)\Big]\Bigg] = -\frac{ig}{\hbar}\int \dif^2 \rp \Tr_\phi\Bigg[\Big[\psi^{\dagger}(\mathbf{r},t) \phi^{\dagger}(\mathbf{r},t)  \phi(\mathbf{r},t) \psi(\mathbf{r},t), \rho_\psi\otimes\rho_\phi\Big]\Bigg].
\end{equation}
Using the cyclic property of the trace we can simplify the above expression to 
\begin{equation}
    \Tr_{\phi}\Bigg[-\frac{i}{\hbar}\Big[H_{int}(t),\rho_I(0)\Big]\Bigg] = -\frac{ig}{\hbar}\int \dif^2\rp \Tr_\phi\Big[\phi^{\dagger}(\mathbf{r},t)  \phi(\mathbf{r},t)\rho_\phi\Big]\otimes\Big[\psi^{\dagger}(\mathbf{r},t)  \psi(\mathbf{r},t), \rho_\psi\Big].
\end{equation}
We recall from the definition of $H_{int}(t)$ above that $\phi^{\dagger}(\mathbf{r},t)  \phi(\mathbf{r},t)=e^{iH_\phi t/\hbar}\phi^{\dagger}(\mathbf{r})  \phi(\mathbf{r})e^{-iH_\phi t/\hbar}$ and again using the cyclic property of the trace we can write the trace term as,
\begin{equation}
Tr_\phi\Big[\phi^{\dagger}(\mathbf{r},t)  \phi(\mathbf{r},t)\rho_\phi\Big]=Tr_\phi\Big[\phi^{\dagger}(\mathbf{r})  \phi(\mathbf{r})e^{-iH_\phi t/\hbar}\rho_\phi e^{i H_\phi t/\hbar}\Big]. 
\end{equation}
The time dependence in the above trace drops out because $\Big[H_\phi,\rho_\phi\Big]=0$. Now let us make an assumption that the electron reservoir remains in the thermal equilibrium at all times, which would imply that
\begin{equation}
Tr_\phi\Big[\phi^{\dagger}(\mathbf{r})  \phi(\mathbf{r})\rho_\phi\Big]=n_\phi(\mathbf{r}), 
\end{equation}
where $n_\phi(\mathbf{r})$ is the local density of the  reservoir. Putting this back in Equation (12), 
\begin{equation}
    \Tr_{\phi}\Bigg[-\frac{i}{\hbar}\Big[H_{int}(t),\rho_I(0)\Big]\Bigg] = -\frac{ig}{\hbar}\int \dif^2\rp n_\phi(\mathbf{r})\Big[\psi^{\dagger}(\mathbf{r},t)  \psi(\mathbf{r},t), \rho_\psi\Big].
\end{equation}
We have already absorbed this term in the definition of $H_\psi$ as given in \eqref{eqn:Hpsi&phi}. Let us now simplify the commutator in the second term in \eqref{eqn:MasterEqn}
\begin{equation}\label{eqn:comex}
\begin{aligned}
    \Bigg[H_{int}(t),\Big[H_{int}(t'),\rho_I(t')\Big]\Bigg] = {} & H_{int}(t)H_{int}(t')\rho_I(t') - H_{int}(t)\rho_I(t')H_{int}(t') \\
    & - H_{int}(t')\rho_I(t')H_{int}(t) + \rho_I(t')H_{int}(t')H_{int}(t).
\end{aligned}
\end{equation}
\begin{equation}
H_{int}(t)H_{int}(t')\rho_I(t') = g^2\int \dif^2\rp \int \dif^2\rp' \psi^{\dagger}(1) \phi^{\dagger}(1)  \phi(1) \psi(1) \psi^{\dagger}(2) \phi^{\dagger}(2)  \phi(2) \psi(2) \rho_I(t')
\end{equation}
where we have used a short hand notation, 1 = $(\mathbf{r},t)$ and 2 = $(\mathbf{r}',t')$. Taking partial trace on this term gives,
\begin{equation}\label{eqn:simp1}
    g^2\int \dif^2\rp \int \dif^2\rp' \psi^{\dagger}(1) \psi(1) \psi^{\dagger}(2) \psi(2) \Tr_\phi \Big[\phi^{\dagger}(1)\phi(1) \phi^{\dagger}(2)  \phi(2)\rho_I(t')\Big].
\end{equation}
Similarly the other terms in \eqref{eqn:comex} could be simplified as,
\begin{equation}
- H_{int}(t)\rho_I(t')H_{int}(t') = -g^2\int \dif^2\rp \int \dif^2\rp' \psi^{\dagger}(1) \phi^{\dagger}(1)  \phi(1) \psi(1) \rho_I(t') \psi^{\dagger}(2) \phi^{\dagger}(2)  \phi(2) \psi(2) 
\end{equation}
Taking partial trace on this term gives,
\begin{equation}\label{eqn:simp2}
    -g^2\int \dif^2\rp \int \dif^2\rp' \psi^{\dagger}(1) \psi(1)  \Tr_\phi \Big[\phi^{\dagger}(1)\phi(1) \rho_I(t') \phi^{\dagger}(2)  \phi(2)\Big]\psi^{\dagger}(2) \psi(2).
\end{equation}
The third term $- H_{int}(t')\rho_I(t')H_{int}(t)$ simplifies to
\begin{equation}
- H_{int}(t')\rho_I(t')H_{int}(t) = -g^2\int \dif^2\rp \int \dif^2\rp' \psi^{\dagger}(2) \phi^{\dagger}(2)  \phi(2) \psi(2) \rho_I(t') \psi^{\dagger}(1) \phi^{\dagger}(1)  \phi(1) \psi(1). 
\end{equation}
Taking partial trace we find,
\begin{equation}\label{eqn:simp3}
    -g^2\int \dif^2\rp \int \dif^2\rp' \psi^{\dagger}(2) \psi(2)  \Tr_\phi \Big[\phi^{\dagger}(2)\phi(2) \rho_I(t') \phi^{\dagger}(1)  \phi(1)\Big]\psi^{\dagger}(1) \psi(1).
\end{equation}
Finally, the fourth term is
\begin{equation}
\rho_I(t')H_{int}(t')H_{int}(t) = g^2\int \dif^2\rp \int \dif^2\rp' \rho_I(t') \psi^{\dagger}(2) \phi^{\dagger}(2)  \phi(2) \psi(2)  \psi^{\dagger}(1) \phi^{\dagger}(1)  \phi(1) \psi(1)  
\end{equation}
which after taking trace gives
\begin{equation}\label{eqn:simp4}
    g^2\int \dif^2\rp \int \dif^2\rp'  \Tr_\phi \Big[\rho_I(t') \phi^{\dagger}(2)\phi(2) \phi^{\dagger}(1)  \phi(1)\Big]\psi^{\dagger}(2) \psi(2) \psi^{\dagger}(1) \psi(1).
\end{equation}
Let us introduce another short hand notation for the trace 
\begin{equation}
\Tr_\phi(12) = \Tr_\phi \Big[\phi^{\dagger}(1)\phi(1) \phi^{\dagger}(2)  \phi(2)\rho_I(t')\Big],
\end{equation}
\begin{equation}
\Tr_\phi(21) = \Tr_\phi \Big[\phi^{\dagger}(2)\phi(2) \phi^{\dagger}(1)  \phi(1)\rho_I(t')\Big].
\end{equation}
Using the cyclic property of the trace we can combine terms \eqref{eqn:simp1} and \eqref{eqn:simp3} into
\begin{equation}
g^2\int \dif^2\rp \int \dif^2\rp' \Big[\psi^{\dagger}(1) \psi(1), \psi^{\dagger}(2) \psi(2)\Tr_\phi(12)\Big].
\end{equation}
Similarly combining terms \eqref{eqn:simp2} and \eqref{eqn:simp4} gives
\begin{equation}
g^2\int \dif^2\rp \int \dif^2\rp' \Big[\Tr_\phi(21)\psi^{\dagger}(2) \psi(2), \psi^{\dagger}(1) \psi(1)\Big].
\end{equation}
To make further progress we'll make assumptions about $\rho_I(t')$. First we assume that the interactions are weak so that correlations between the reservoir and the polaritons are small even at long time scales. This motivates us to approximate $\rho_I(t')\approx \rho^I_\psi(t')\otimes\rho_\phi$. Secondly we'll assume that the energy separation scale between the electrons and the polaritons is large. This will result in a much slower evolution of $\rho^I_\psi(t')$ allowing us to replace $\rho^I_\psi(t')\rightarrow\rho^I_\psi(t)$ in the time integral. This approximation is also known as the Markov approximation because it says that the evolution of the density matrix depends on the instantaneous state of the system and makes no reference to the past. We'll introduce another notation for the trace in terms of calculating thermal reservoir average, $\Tr_\phi[\rho_\phi \ .\ ] = \langle \ .\ \rangle$. Therefore the above expressions will approximate to,
\begin{equation}\label{eqn:g2_1}
\begin{aligned}
g^2\int \dif^2\rp \int \dif^2\rp' \Big[\psi^{\dagger}(1) \psi(1), \psi^{\dagger}(2) \psi(2) \rho^I_\psi(t)\Big]\otimes\langle\phi^{\dagger}(1)\phi(1)\phi^{\dagger}(2)\phi(2)\rangle & \\
\end{aligned}
\end{equation}
\begin{equation}\label{eqn:g2_2}
\begin{aligned}
g^2\int \dif^2\rp \int \dif^2\rp' \Big[\rho^I_\psi(t)\psi^{\dagger}(2) \psi(2), \psi^{\dagger}(1) \psi(1)\Big]\otimes\langle\phi^{\dagger}(2)\phi(2)\phi^{\dagger}(1)\phi(1)\rangle & \\
\end{aligned}
\end{equation}
Let us now simplify the temporal part of the reservoir correlation function
\begin{equation}\label{eqn:g2_3}
\begin{aligned}
\langle\phi^{\dagger}(2)\phi(2)\phi^{\dagger}(1)\phi(1)\rangle = & \langle e^{iH_\phi t'/\hbar}\phi^{\dagger}(\mathbf{r}')\phi(\mathbf{r}')e^{-iH_\phi t'/\hbar}e^{iH_\phi t/\hbar}\phi^{\dagger}(\mathbf{r})\phi(\mathbf{r})e^{-iH_\phi t/\hbar}\rangle \\
& = \langle \phi^{\dagger}(\mathbf{r}')\phi(\mathbf{r}')e^{-iH_\phi t'/\hbar}e^{iH_\phi t/\hbar}\phi^{\dagger}(\mathbf{r})\phi(\mathbf{r})e^{-iH_\phi t/\hbar}e^{iH_\phi t'/\hbar}\rangle \\
& = \langle \phi^{\dagger}(\mathbf{r}')\phi(\mathbf{r}')e^{iH_\phi (t-t')/\hbar}\phi^{\dagger}(\mathbf{r})\phi(\mathbf{r})e^{-iH_\phi (t-t')/\hbar}\rangle \\
& = \langle \phi^{\dagger}(\mathbf{r}',0)\phi(\mathbf{r}',0)\phi^{\dagger}(\mathbf{r},\tau)\phi(\mathbf{r},\tau)\rangle,
\end{aligned}
\end{equation}
where we have defined $\tau = t-t'$. Repeating the same steps we have
\begin{equation}\label{eqn:g2_4}
\langle\phi^{\dagger}(1)\phi(1)\phi^{\dagger}(2)\phi(2)\rangle = \langle \phi^{\dagger}(\mathbf{r},\tau)\phi(\mathbf{r},\tau)\phi^{\dagger}(\mathbf{r}',0)\phi(\mathbf{r}',0)\rangle.
\end{equation}
Also the commutators in the expressions \eqref{eqn:g2_1} and \eqref{eqn:g2_2} would be,
\begin{equation}\label{eqn:supOp1}
\begin{aligned}
 \Big[\psi^{\dagger}(1) \psi(1), \psi^{\dagger}(2) \psi(2) \rho^I_\psi(t)\Big] & =\Big[\psi^{\dagger}(\mathbf{r},t) \psi(\mathbf{r},t), \psi^{\dagger}(\mathbf{r}',t-\tau) \psi(\mathbf{r}',t-\tau) \rho^I_\psi(t)\Big] \\
& =\Big[\hat{n}_\psi(\mathbf{r},t), \hat{n}_\psi(\mathbf{r}',t-\tau)\rho^I_\psi(t)\Big]
\end{aligned}
\end{equation}
\begin{equation}\label{eqn:supOp2}
\begin{aligned}
\Big[\rho^I_\psi(t)\psi^{\dagger}(2) \psi(2), \psi^{\dagger}(1) \psi(1) \Big] & =\Big[\rho^I_\psi(t)\psi^{\dagger}(\mathbf{r}',t-\tau) \psi(\mathbf{r}',t-\tau), \psi^{\dagger}(\mathbf{r},t) \psi(\mathbf{r},t) \Big] \\
& =\Big[\rho^I_\psi(t)\hat{n}_\psi(\mathbf{r}',t-\tau), \hat{n}_\psi(\mathbf{r},t) \Big]
\end{aligned}
\end{equation}
Going back to the reservoir correlation functions we'll apply Hartree Fock factorization to express the four field correlation function as a product of two field correlators. Keeping in mind that the anti-commutation relation results in a negative sign for odd number of field exchanges, we see that the only relevant factorization involves even exchanges. We'll drop anomalous paired correlators like $\langle\phi^\dagger\phi^\dagger\rangle$ and $\langle\phi\phi\rangle$ because they don't conserve particle number. Also equal time correlators like $\langle\phi^\dagger\phi\rangle$ will be dropped because they are time independent, which will lead to incorrect energy conservation relation as seen after doing the integral w.r.t. $\tau$. Therefore correlators \eqref{eqn:g2_3} and \eqref{eqn:g2_4} simplify to 
\begin{equation}\label{eqn:corr1}
    \langle \phi^{\dagger}(\mathbf{r}',0)\phi(\mathbf{r}',0)\phi^{\dagger}(\mathbf{r},\tau)\phi(\mathbf{r},\tau)\rangle = \langle\phi^{\dagger}(\mathbf{r}',0)\phi(\mathbf{r},\tau)\rangle\langle\phi(\mathbf{r}',0)\phi^{\dagger}(\mathbf{r},\tau)\rangle
\end{equation}
\begin{equation}\label{eqn:corr2}
\langle \phi^{\dagger}(\mathbf{r},\tau)\phi(\mathbf{r},\tau)\phi^{\dagger}(\mathbf{r}',0)\phi(\mathbf{r}',0)\rangle = \langle\phi^{\dagger}(\mathbf{r},\tau)\phi(\mathbf{r}',0)\rangle\langle\phi(\mathbf{r},\tau)\phi^{\dagger}(\mathbf{r}',0)\rangle
\end{equation}
We can evaluate the above correlation functions by introducing the field operator for the electrons,
\begin{equation}
    \phi(\vec{r},t) = \frac{1}{L}\sum_k e^{i(\vec{k}.\vec{r} - \omega_k t)} b_{\vec{k}}. 
\end{equation}
The plane wave basis is chosen since the electrons are assumed to be propagating freely in the quantum well. The density of the free electrons are assumed to be low such that we can ignore the repulsive interactions between them.
Thermal expectation value for the operator $b^\dagger_{\vec{k}} b_{\vec{k}}$ is given by,
\begin{equation}
    \langle b^\dagger_{\vec{k}} b_{\vec{k}} \rangle = \sum_{\vec{k}} n_{\vec{k}} = \frac{L^2}{4\pi^2}\int \dif^2 \kp \frac{1}{e^{(E_\phi(\vec{k})-\mu)/k_B T}+1}
\end{equation}
where $E_\phi(\vec{k})/\hbar = \omega_{\vec{k}} = \hbar \vec{k}^2/2m_\phi$. Calculating products of the field operators at different space and time coordinates is straightforward,
\begin{equation}
    \phi^{\dagger}(\mathbf{r}',t')\phi(\mathbf{r},t) = \frac{1}{L^2}\sum_{\vec{k},\vec{k}'} b^\dagger_{\vec{k}'}b_{\vec{k}} e^{-i \vec{k}'\cdot\vec{r}' + i\vec{k}\cdot\vec{r} - i\omega_{\vec{k}} t + i \omega_{\vec{k}'} t'}.
\end{equation}
Using these we can now look at the correlators,
\begin{equation}
\begin{aligned}
    \langle\phi^{\dagger}(\mathbf{r}',0)\phi(\mathbf{r},\tau)\rangle & = \ \frac{1}{L^2}\sum_{\vec{k},\vec{k}'}\langle b^\dagger_{\vec{k}'}b_{\vec{k}}\rangle e^{-i \vec{k}'.\vec{r}' + i\vec{k}.\vec{r} - i\omega_{\vec{k}} \tau} \\
    & =  \frac{1}{L^2}\sum_{\vec{k},\vec{k}'}n_{\vec{k}} \delta_{\vec{k},\vec{k}'} e^{-i \vec{k}'.\vec{r}' + i\vec{k}.\vec{r} - i\omega_{\vec{k}} \tau} \\
    & =  \frac{1}{4\pi^2}\int \dif^2 \kp n_{\vec{k}} e^{i\vec{k}.(\vec{r}-\vec{r}')-iE_\phi\tau/\hbar}.
\end{aligned}
\end{equation}
\begin{equation}
\begin{aligned}
    \langle\phi(\mathbf{r}',0)\phi^{\dagger}(\mathbf{r},\tau)\rangle & = \ \frac{1}{L^2}\sum_{\vec{k},\vec{k}'}\langle b_{\vec{k}'}b^\dagger_{\vec{k}}\rangle e^{i \vec{k}'.\vec{r}' - i\vec{k}.\vec{r} + i\omega_{\vec{k}} \tau} \\
    & =  \frac{1}{L^2}\sum_{\vec{k},\vec{k}'}\Big(1-n_{\vec{k}}\Big) \delta_{k,k'} e^{i \vec{k}'.\vec{r}' - i\vec{k}.\vec{r} + i\omega_{\vec{k}} \tau} \\
    & =  \frac{1}{4\pi^2}\int \dif^2 \kp \Big(1-n_{\vec{k}}\Big) e^{-i\vec{k}.(\vec{r}-\vec{r}')+iE_\phi\tau/\hbar}.
\end{aligned}
\end{equation}
\begin{equation}
\begin{aligned}
    \langle\phi^{\dagger}(\mathbf{r},\tau)\phi(\mathbf{r}',0)\rangle & = \ \frac{1}{L^2}\sum_{\vec{k},\vec{k}'}\langle b^\dagger_{\vec{k}}b_{k'}\rangle e^{i \vec{k}'.\vec{r}' - i\vec{k}.\vec{r} + i\omega_{\vec{k}} \tau} \\
    & =  \frac{1}{L^2}\sum_{\vec{k},\vec{k}'}n_{\vec{k}} \delta_{\vec{k},\vec{k}'} e^{i \vec{k}'.\vec{r}' - i\vec{k}.\vec{r} + i\omega_{\vec{k}} \tau} \\
    & =  \frac{1}{4\pi^2}\int \dif^2\kp n_{\vec{k}} e^{-i\vec{k}.(\vec{r}-\vec{r}')+iE_\phi\tau/\hbar}.
\end{aligned}
\end{equation}
\begin{equation}
\begin{aligned}
    \langle\phi(\mathbf{r},\tau)\phi^{\dagger}(\mathbf{r}',0)\rangle & = \ \frac{1}{L^2}\sum_{\vec{k},\vec{k}'}\langle b_{\vec{k}} b^\dagger_{\vec{k}'}\rangle e^{-i \vec{k}'.\vec{r}' + i\vec{k}.\vec{r} - i\omega_{\vec{k}} \tau} \\
    & =  \frac{1}{L^2}\sum_{\vec{k},\vec{k}'}\Big(1-n_{\vec{k}}\Big) \delta_{\vec{k},\vec{k}'} e^{-i \vec{k}'.\vec{r}' + i\vec{k}.\vec{r} - i\omega_k \tau} \\
    & =  \frac{1}{4\pi^2}\int \dif^2\kp \Big(1-n_{\vec{k}}\Big) e^{i\vec{k}.(\vec{r}-\vec{r}')-iE_\phi\tau/\hbar}.
\end{aligned}
\end{equation}
With the definition $n_{\vec{k}}=F(\mathbf{k})=1/(e^{(E_\phi(\vec{k})-\mu)/k_B T}+1)$ and using the above results we can write \eqref{eqn:corr1} as 
\begin{equation}\label{eqn:prod1}
    \frac{1}{(2\pi)^4}\int \dif^2\kp_1 \int \dif^2\kp_2 F(\mathbf{k_1})\Big(1-F(\mathbf{k_2})\Big) e^{i(\mathbf{k_1}-\mathbf{k_2}).(\mathbf{r}-\mathbf{r}')-i(E_\phi(\mathbf{k_1})-E_\phi(\mathbf{k_2}))\tau/\hbar},
\end{equation}
and \eqref{eqn:corr2} as 
\begin{equation}\label{eqn:prod2}
    \frac{1}{(2\pi)^4}\int \dif^2\kp_1 \int \dif^2\kp_2 F(\mathbf{k_1})\Big(1-F(\mathbf{k_2})\Big) e^{-i(\mathbf{k_1}-\mathbf{k_2}).(\mathbf{r}-\mathbf{r}')+i(E_\phi(\mathbf{k_1})-E_\phi(\mathbf{k_2}))\tau/\hbar}.
\end{equation}
In expressions \eqref{eqn:supOp1} and \eqref{eqn:supOp2} we could define an operator $\hat{L}_\psi$ whose action is defined through $\hat{L}_\psi\hat{n}_\psi(\mathbf{r}')= [\hat{n}_\psi(\mathbf{r}'),H_\psi]$, which leads to $\hat{n}_\psi(\mathbf{r}',t)=e^{-i\hat{L}_\psi t/\hbar}\hat{n}_\psi(\mathbf{r}')$. We will make a change of variable for the integral over $t'$ in equation \eqref{eqn:MasterEqn},
\begin{equation}
    \int_0^t \dt' \rightarrow  \int_0^t d\mathrm{\tau}
\end{equation}
and also set the upper limit of the integral to $\infty$. This is done with a presumption that if we wait long enough we can recover energy and momentum conserving scattering between the coherent polaritons and the reservoir. This is usually true if $t\gg$ correlation time of the reservoir. Doing the $\tau$ integral one obtains for product of the expressions \eqref{eqn:supOp1} and \eqref{eqn:prod2},  
\small
\begin{equation}
 \hbar \pi \delta\left(\hat{L}_\psi+E_\phi(\mathbf{k_1})-E_\phi(\mathbf{k_2})\right)
  \end{equation}
and for \eqref{eqn:supOp2} and \eqref{eqn:prod1}
\begin{equation}
   \hbar \pi \delta\left(\hat{L}_\psi-E_\phi(\mathbf{k_1})+E_\phi(\mathbf{k_2})\right).
\end{equation}
We have now calculated everything we need to put back in equation \eqref{eqn:MasterEqn} to obtain \eqref{eqn:densityMatrix1}.
\end{widetext}
\section{Weyl-Wigner transformations}\label{appendix:Appendix_transform}
Strings of quantum operators appearing in Equation (\ref{eqn:densityMatrix2}) are mapped to classical variables using Wigner-Weyl transformations. Readers are directed to references \cite{gardiner2004quantum,polkovnikov2010phase} for an introduction to semi-classical transformations which are used in studying quantum dynamics using phase space methods. In this section we simply provide the transformation rules for the three types of terms which are encountered while expanding the commutator in Equation (\ref{eqn:densityMatrix2}).
\begin{widetext}
\begin{equation}
\begin{aligned}
    & \psi^\dagger\psi\psi^\dagger\psi\rho\longrightarrow\bigg(\psi^*-\frac{1}{2}\delta_\psi\bigg)\bigg(\psi+\frac{1}{2}\delta_{\psi^*}\bigg)\bigg(\psi^*-\frac{1}{2}\delta_\psi\bigg)\bigg(\psi+\frac{1}{2}\delta_{\psi^*}\bigg)W(\psi,\psi^*)\\
    & \psi^\dagger\psi\rho\psi^\dagger\psi\longrightarrow\bigg(\psi^*-\frac{1}{2}\delta_\psi\bigg)\bigg(\psi+\frac{1}{2}\delta_{\psi^*}\bigg)\bigg(\psi-\frac{1}{2}\delta_{\psi^*}\bigg)\bigg(\psi^*+\frac{1}{2}\delta_\psi\bigg)W(\psi,\psi^*)\\
    & \rho\psi^\dagger\psi\psi^\dagger\psi\longrightarrow\bigg(\psi-\frac{1}{2}\delta_{\psi^*}\bigg)\bigg(\psi^*+\frac{1}{2}\delta_\psi\bigg)\bigg(\psi-\frac{1}{2}\delta_{\psi^*}\bigg)\bigg(\psi^*+\frac{1}{2}\delta_\psi\bigg)W(\psi,\psi^*).\\
    \end{aligned}
\end{equation}

\end{widetext}

\section{Calculation of the scattering amplitude $M(\mathbf{r}-\mathbf{r'})$}\label{appendix:Appendix_scattering}	
In this section we consider a general case of the electron reservoir drifting with a momentum $\hbar \mathbf{k_0}$, such that the energy of the electrons are now given by $E(\mathbf{k}) = \hbar^2 (\mathbf{k}-\mathbf{k_0})^2/2m_\phi$. We evaluate the scattering amplitude $M(\mathbf{r}-\mathbf{r'})=M(\mathbf{v})$ by defining the Fourier transform $\tilde{M}(\mathbf{q})$,
\begin{widetext}
\begin{equation}\label{eqn:scatteringMatrix3}
\begin{aligned}
    \tilde{M}(\mathbf{q}) & = \frac{1}{(2\pi)^2}\int \dif^2\vp  M(\mathbf{v}) e^{-i\mathbf{q}\cdot\mathbf{v}}\\
    & = \frac{1}{(2\pi)^2}\int \dif^2\vp  \dif^2\kp_1 \dif^2\kp_2 F(\boldsymbol{k_1})\Big(1-F(\boldsymbol{k_2})\Big)\delta\left(E_\phi(\boldsymbol{k_1})-E_\phi(\boldsymbol{k_2})\right)e^{i(\boldsymbol{k_1}-\boldsymbol{k_2}-\mathbf{q})\cdot\mathbf{v}} \\
    & = \int  \dif^2\kp_1 \dif^2\kp_2 F(\boldsymbol{k_1})\Big(1-F(\boldsymbol{k_2})\Big)\delta\left(E_\phi(\boldsymbol{k_1})-E_\phi(\boldsymbol{k_2})\right)\delta\big(\boldsymbol{k_1}-\boldsymbol{k_2}-\mathbf{q}\big)\\
    & = \int  \dif^2\kp_1 F(\boldsymbol{k_1})\Big(1-F(\boldsymbol{k_1}-\mathbf{q})\Big)\delta\left(E_\phi(\boldsymbol{k_1})-E_\phi(\boldsymbol{k_1}-\mathbf{q})\right)\\
    & = \frac{m_\phi}{\hbar^2}\int  \dif^2\kp_1 F(\boldsymbol{k_1})\Big(1-F(\boldsymbol{k_1}-\mathbf{q})\Big)\delta\left(\mathbf{k_1}\cdot\mathbf{q}-\frac{\mathbf{q}^2}{2}-\mathbf{k_0}\cdot\mathbf{q}\right)\\
    & = \frac{m_\phi}{\hbar^2}\int  \dif^2\kp_1 F(\boldsymbol{k_1})\Big(1-F(\boldsymbol{k_1})\Big)\delta\left(\mathbf{k_1}\cdot\mathbf{q}-\frac{\mathbf{q}^2}{2}-\mathbf{k_0}\cdot\mathbf{q}\right)\\
    & = \frac{m_\phi}{\hbar^2}\int  \dif^2\kp_1 F(\boldsymbol{k_1})\Big(1-F(\boldsymbol{k_1})\Big)\frac{1}{|\mathbf{q}|}\delta\left(\mathbf{k_1}\cdot\hat{\mathbf{q}}-\frac{|\mathbf{q}|}{2}-\mathbf{k_0}\cdot\hat{\mathbf{q}}\right)\\
    & = \frac{m_\phi}{\hbar^2}\int  \dif\kp_1^{||}\dif\kp_1^{\bot} F(\boldsymbol{k_1^{\parallel}}+\mathbf{k_1^{\bot}})\Big(1-F(\boldsymbol{k_1^{\parallel}}+\mathbf{k_1^{\bot}})\Big)\frac{1}{|\mathbf{q}|}\delta\left(k_1^{||}-\frac{|\mathbf{q}|}{2}-k_0^{||}\right)\\
    & = \frac{m_\phi}{\hbar^2}\int  \dif\kp_1^{\bot} F\Big(\frac{\mathbf{q}}{2}+\mathbf{k_0^{||}}+\mathbf{k_1^{\bot}}\Big)\Big(1-F\Big(\frac{\mathbf{q}}{2}+\mathbf{k_0^{||}}+\mathbf{k_1^{\bot}}\Big)\Big)\frac{1}{|\mathbf{q}|}
\end{aligned}
\end{equation}
We consider the high temperature and low electron density limit in which $F_{FD}(1-F_{FD})\rightarrow F_{MB}$, where subscripts stand for Fermi-Dirac and Maxwell-Boltzmann distributions respectively. $F_{MB}$ is given by $e^{\beta\mu} e^{-\beta\hbar^2q^2/8m_\phi}e^{-\beta\hbar^2(k_1^{\bot}-k_0^{\bot})^2/2m_\phi}$. Integrating w.r.t. $k_1^{\bot}$ we get  
\begin{equation}
    \tilde{M}(\mathbf{q}) = \frac{m_\phi^{3/2}}{\hbar^3}\sqrt{\frac{2\pi}{\beta}}e^{\beta\mu}\frac{e^{-\beta \hbar^2\mathbf{q}^2/8m_\phi}}{|\mathbf{q}|}.
\end{equation}
\end{widetext}

\section{Drag potential in one-dimension}\label{appendix:Appendix_drag1D}	
Using $\psi(x,y)= \sqrt{(2/w)}\cos{(\pi y/w)} \psi(x)$ we can write $\mathbf{J}(\mathbf{r'})$ as
\begin{equation}
    \mathbf{J}(\mathbf{r'}) = \frac{2}{w}\cos^2\big(\pi y'/w\big)\Big(\mathbf{v}_{rel}|\psi(x')|^2 +\mathbf{j}(x')\Big).
\end{equation}
$\mathbf{j}(x')$ and $\mathbf{v}_{rel}$ has non- zero components only along $\pm$ x-direction. Therefore, $\mathbf{J}(\mathbf{r'})$ is a vector pointing along $\pm$ x-direction. Divergence of this current gives
\begin{equation}
\begin{aligned}
    \mathbf{\nabla'}\cdot\mathbf{J}(\mathbf{r'}) = & \frac{2}{w}\cos^2\big(\pi y'/w\big)\Big(|\mathbf{v}_{rel}|(\mathbf{\hat{v}}_{rel}\cdot\mathbf{\hat{x}'})\partial_{x'}|\psi(x')|^2 \\
    & \ \ +\partial_{x'}j(x')\Big).
\end{aligned}
\end{equation}
Substituting for the divergence of current and the inverse Fourier transform of $M(\mathbf{r}-\mathbf{r'})$ in the expression for $V_\epsilon(\mathbf{r})$ in \eqref{eqn:densityMatrix3}
\begin{widetext}
\begin{equation}
\begin{aligned}
    V_\epsilon(\mathbf{r})& = -\frac{\hbar\beta g^2}{32\pi^3w}\int \dif^2\rp' \dif^2\q e^{i\mathbf{q}\cdot\mathbf{r}}\tilde{M}(\mathbf{q})e^{-i\mathbf{q}\cdot\mathbf{r'}}\bigg(1+\cos\big(2\pi y'/w\big)\bigg)\Big(|\mathbf{v}_{rel}|(\mathbf{\hat{v}}_{rel}\cdot\mathbf{\hat{x}'})\partial_{x'}|\psi(x')|^2  +\partial_{x'}j(x')\Big), \\
    & = -\frac{2\pi\hbar\beta g^2}{32\pi^3w}\int \dif \x' \dif\q_x e^{iq_x x}\bigg(\tilde{M}(q_x,0)+\frac{e^{i2\pi y/w}}{2}\tilde{M}\big(q_x,2\pi/w\big)+\frac{e^{-i2\pi y/w}}{2}\tilde{M}\big(q_x,-2\pi/w\big)\bigg)\\
    &\ \ \ \ \ \ \ \ \ \ \ \ \ \ \ \ \ \ \ \ \ \ \ \ \ \ \ \ \  \Big(|\mathbf{v}_{rel}|(\mathbf{\hat{v}}_{rel}\cdot\mathbf{\hat{x}'})\partial_{x'}|\psi(x')|^2  +\partial_{x'}j(x')\Big)e^{-iq_x x'}, \\
    & = -\frac{2\pi\hbar\beta g^2}{16\pi^2w}\int \dif\q_x e^{iq_x x}\bigg(\tilde{M}(q_x,0)+\frac{e^{i2\pi y/w}}{2}\tilde{M}\big(q_x,2\pi/w\big)+\frac{e^{-i2\pi y/w}}{2}\tilde{M}\big(q_x,-2\pi/w\big)\bigg)\mathcal{J}(q_x).
\end{aligned}
\end{equation}
$\mathcal{J}(q_x)$ is defined as the Fourier transform of $\big(|\mathbf{v}_{rel}|(\mathbf{\hat{v}}_{rel}\cdot\mathbf{\hat{x}'})\partial_{x'}|\psi(x')|^2  +\partial_{x'}j(x')\big)$. Finally, to obtain $V_\epsilon(x)$ we simply multiply $V_\epsilon(\mathbf{r})$ with the transverse probability of finding the particles and integrate over the transverse direction and is given by,
\begin{equation}
     \begin{aligned}
        V_\epsilon(x) & = \frac{2}{w}\int_{-w/2}^{w/2} \dy \cos^2\big(\pi y/w\big) V_\epsilon(\mathbf{r}), \\
        &  = -\frac{2\pi\hbar\beta g^2}{16\pi^2w}\int \dif\q_x e^{iq_x x}\bigg(\tilde{M}(q_x,0)+\frac{1}{4}\tilde{M}\big(q_x,2\pi/w\big)+\frac{1}{4}\tilde{M}\big(q_x,-2\pi/w\big)\bigg)\mathcal{J}(q_x). 
    \end{aligned}
\end{equation}
\end{widetext}

\end{document}